\documentclass[11pt]{llncs}
\usepackage{times}

\usepackage{amsmath}
\usepackage{llncsdoc}

\usepackage{amssymb}

\usepackage{graphicx}

\usepackage[]{algorithm2e}
\usepackage{listings}
\usepackage{caption}

\usepackage{fancyhdr}
\fancypagestyle{plain}{
\fancyhf{} 
\fancyfoot[C]{\thepage} 
  
}

\usepackage[margin=1.181in]{geometry}
\usepackage{times}
\usepackage[affil-it]{authblk}

\begin{document}
\newcommand*{\myfont}{\fontfamily{mdugm}\selectfont}
\newcommand*{\absfont}{\fontfamily{times}\fontsize{10}{10}\selectfont}
\newcommand*{\kwords}{\fontfamily{times}\fontsize{10}{10}\selectfont}

\begin{myfont}
\title{The Bases of Association Rules of High Confidence}
\end{myfont}

\renewcommand\Authfont{\fontsize{13}{13}\selectfont}

\author{Oren Segal\inst{1} \and Justin Cabot-Miller\inst{2} \and Kira Adaricheva\inst{2} \and J.B.Nation \inst{3}  \and Anuar Sharafudinov \inst{4}}
\institute{Department of Computer Science, Hofstra University, Hempstead, NY 11549\\ Oren.Segal@hofstra.edu \and Department of Mathematics, Hofstra University, Hempstead, NY 11549\\ JCabotMiller1@pride.hofstra.edu and Kira.Adaricheva@hofstra.edu \and Department of Mathematics, University of Hawaii, Honolulu HI 11823\\ JB@math.hawaii.edu \and AILabs, Astana, 010000 Kazakhstan\\AnuarSh@ailabs.kz}

\maketitle


\begin{abstract}
\begin{absfont}
We develop a new approach for distributed computing of the association rules of high confidence in a binary table. It is derived from the $D$-basis algorithm \cite{AN15}, which is performed on multiple sub-tables of a table given by removing several rows at a time. The set of rules is then aggregated using the same approach as the $D$-basis is retrieved from a larger set of implications. This allows to obtain a basis of association rules of high confidence, which can be used for ranking all attributes of the table with respect to a given fixed attribute using the \emph{relevance} parameter introduced in \cite{ANO15}. This paper focuses on the technical implementation of the new algorithm. Some testing results are performed on transaction data and medical data. 
\end{absfont}
\end{abstract}



\pagestyle{myheadings}
\thispagestyle{plain}
\markboth{The bases of association rules}{O. Segal, J. Cabot-Miller, K. Adaricheva, J.B.Nation, A. Sharafudinov}

\section{Introduction}
\pagestyle{empty}



In data mining, the retrieval and sorting of association rules is a research problem of considerable interest. Association rules uncover the relationships between the attributes of a set of objects recorded in a binary table. This, for example, can be a transaction table, where the objects are sale transactions and the attributes are groups of products: position $(r,c)$ in the table, in row $r$ and column $c$, is marked by $1$ if the transaction $r$ includes a product from that group $c$, otherwise, it is marked by $0$. Association rule $X\rightarrow b$ in the table means that the entire set of transactions shows the tendency that whenever a transaction includes products from all groups in set $X$ (i.e., there are $1$s in all columns from $X$), then some product in group $b$ will appear as well. The confidence of such a rule is measured by a portion of all transactions that include $b$ among all transactions that have products from $X$.

In the world of transaction data, a rule $X\rightarrow b$ with confidence of $0.1$ might demonstrate that  $b$ \emph{sells together} with all products in group $X$.

There is an immense effort in the data mining community to develop reliable tools for the discovery of meaningful association rules. However, the hurdles encountered while developing such solutions are numerous.
The benchmark algorithms, such as  \emph{Apriori} in Agrawal et al.~\cite{A96}, have time complexity that is exponential in regards to the size of the input table. Moreover, the number of association rules is staggering, and thus analyzing them requires further tools to obtain a short subset of rules that are significant. There are no strong mathematical results confirming a particular choice of such short subsets, and numerous approaches to the filtering process are described in various publications devoted to the topic. See, for example, Kryszkiewicz \cite{K02} and Balc\'azar \cite{B10}. Recent approaches include constraint-based patterns, preference learning, instant mining and pattern space sampling, which are often interactive methods targeting user's implicit preferences, see \cite{BPBBK16}, \cite{SC09}, \cite{SRPC11}.


One particular subset of association rules, the \emph{implications}, or rules of full confidence, merit particular attention in data mining. They are also at the center of ongoing theoretical research. 
 From a practical point of view, implications are the strongest rules available in a given table because they hold true for any row of the table.

Some types of data present cases where the implications uncovered in the table contain non-essential information because the support of those implications might be very low. The support of association rule $X\rightarrow b$ is the number of rows where ones appear in all columns from $X\cup b$. For example, the transaction data might have only a few implications with small sets $X$, whose support could be a single-digit percentage of all transactions. Mining of transaction data tends to uncover rules of lower confidence but relatively large support, rather than those implications which hold everywhere.

The Apriori algorithm and the concept of generating rules of lower confidence are relevant when considering medical data. The attributes of a table that represent genetic and clinical data of patients (rows) may have a tighter connection than relationships in transaction data, in which case the confidence of relevant association rules could be expected to be high and closer to 1. Implications would serve as the imperfect representation of the laws of nature in this data. At the same time, every data set may contain errors, missing entries and miscalculations. Additionally, some patients may have extraneous conditions affecting the value in the target column (e.g., co-morbidity with an untracked illness). Even if only one row contains such deviations, it may prevent us from discovering important implications that would otherwise hold in the table.


Because errors may exist in the data in small numbers, the type of association rules one would want to discover would be those rules whose confidence is sufficiently close to one. Where "sufficiently close" can be decided on a case-by-case basis.


In this paper, we expand the approach developed in \cite{ANO15} of the extraction of implications and ranking of attributes with respect to a target attribute.


Our goal is not to uncover particular rules and rank them with respect to some measurement. Rather, we want to generate a basis $\Delta$ of association rules which satisfactorily describes dependencies among attributes. We could then use $\Delta$ to rank the importance of attributes with respect to target attribute, $b$. In medical data, $b$ may describe high survival probability of a patient after particular treatment, when other attributes may record physical parameters in the patients. Having large $\Delta$ is not necessarily a bad feature; on the other hand, the optimally small set is desirable. Consider the case where: 

 $X\rightarrow b$, $X\cup c \rightarrow b$ and $X\cup d \rightarrow b$ are in the basis and have high confidence. We may want to keep $A\rightarrow b$ in $\Delta$ and remove the other rules which may unnecessarily inflate the relevance of attributes $c$ and $d$ for $b$. This is because attributes $c, d$ could be completely unrelated to $b$, however only appear because they are not explicitly \emph{blocked} elsewhere.


The paper is organized as follows. We give a short description of the proposed algorithm in section 2. In section 3, we explain how the association rules are used to compute the relevance parameter of one attribute with respect to the other. In section 4, we discuss the technical implementation of the algorithm. We discuss the performance of the algorithm and comparison with existing implementations of Apriori algorithm in section 5. In the final section, section 6, we summarize the future work that we plan to do with our approach.

\section{General flow of the algorithm}

Herein we describe a several stage approach that allows us to compute, beyond just implications, the potentially most valuable association rules whose confidence is rather high, say, $> 0.9$.


At the core of our approach is the connection between a binary table, its implications, and the closure operator defined on the set of columns and associated lattice of closed sets, known as the concept lattice or Galois lattice, see \cite{GW99}.

An algorithm in Adaricheva and Nation \cite{AN15} works to extract the basis of implications of the table, and it is known as the $D$-\emph{basis}, which was introduced in Adaricheva, Nation, and Rand \cite{ANR11}. 
The advantage of this basis is in the possibility to use algorithms for dualization of an associated hypergraph that are known to be sub-exponential in their complexity, see Fredman and Khachiyan \cite{FK96}. The algorithm in \cite{AN15} avoids generating the Galois lattice from the table and only uses the arrow relations, which can be computed in polynomial time, to produce a hypergraph for each requested attribute.  In that way,  the existing code for hypergraph dualization, such as in Murakami and Uno \cite{MU13}, can be borrowed for execution.

The main idea of the current algorithm follows from the observation that
the association rules of high confidence may be computed by removing one or more rows (objects) from the table and computing implications (on attributes) of the shorter table. Upon the program's execution and output, one can record only new implications that were not present in the original data set. 

A new rule may be derived from the shorter table given that one of the removed rows contradicts it. If new rules are present in this table and exhibit high support, or if numerous new rules are found with average support, then the row(s) temporarily removed to form this shorter table are called \emph{blockers} due to their tendency to block the rules that were found with their removal.


We can choose various strategies to identify the set of blockers. Together with several straightforward statistics on new rules found on a shorter table, we are considering several heuristics and ranking systems.
The goal is to identify a set $S$ of rows/objects that are potential blockers.

In the next stage of the suggested procedure we choose $n\leq |S|$ which is a number of rows from $S$ that will be deleted from original table to form a shorter table. With $S$ and $n$ fixed, we can run algorithm $k$ times, where $k$ is specified by a user, and limited by $C(|S|,n)=\frac{n!}{|S|!(|S|-n)!}$, which is a number of combinations to choose $n$ rows from set $S$. If $k < C(|S|,n)$, then the choice can be done randomly, otherwise, the rows can by systematically removed. 


This process of removing sets of rows and re-running the program can be organized in parallel, and all the outputs combined and aggregated following the same procedure as used to aggregate the $D$-basis of implications in original $D$-basis algorithm. 

The final set of rules is guaranteed to have the probability of at least $\frac{N-n}{N}$ that each rules holds in a table, where $N$ is the total number of rows(objects) in the table, and $n$ is the number of deleted rows in each run of the algorithm. If $s$ is the support of some implication $A\rightarrow b$ in a shortened table with $N-n$ rows, then on average the support of $A$ in $n$ deleted rows will be $s\cdot \frac{n}{N-n}$. 

In worst case scenario, i.e. when $b=0$ in all deleted rows, the support of $A\rightarrow b$ on $n$ deleted rows will be $0$. This will give a lower estimate of the confidence of association rule $A\rightarrow b$ in the full table as $\frac{s}{s+s\cdot \frac{n}{N}}=\frac{N}{N+n}$. 

For example, given an original table of 90 rows, the confidence of a rule found as an implication in a sub-table of 80 rows, i.e., after deleting 10 rows, will be, on average, around $\frac{90}{100}=0.9$.

\section{Ranking the attributes of the table with respect to a given fixed attribute}



The algorithm described in the previous section, when we try to identify the \emph{blockers} among the objects/rows of the data, can be interpreted as \emph{unsupervised} learning about the data set. These blockers are then interpreted as outliers. 

In this section we will take a look at \emph{supervised} exploration of the data set, when one of the  parameters is a target column/attribute, and we try to discover other attributes which might be essential for describing the behavior of the target parameter. 

The algorithm in \cite{AN15} allows us to retrieve only those implications $X\rightarrow b$ in the $D$-basis that have a fixed attribute $b$ as a conclusion. This is called a $b$-\emph{sector} of the basis. It is important to notice, for comparison with Apriori algorithm in section 5, that one does not need to obtain the full basis in order to get particular $b$-sector of the basis. 

In our current approach, instead of the set of implications, we obtained the set $\Delta$ of association rules, where we traded the confidence for the higher support of the rules.
We choose a number of shorter tables, as described in algorithm of section 2, and compute $b$-sectors of implications. 
A final part of the algorithm then performs a special trimming of the rules, called an aggregation, only leaving the \emph{strongest} rules with respect to the \emph{binary part} of $\Delta$, which consists of rules of the form $a\rightarrow d$. Thus, in order to obtain the $b$-sector of basis $\Delta$, one needs only the binary part of $\Delta$ and combination of $b$-sectors of implications of shorter tables. The resulting $b$-sector of $\Delta$ after its aggregation will be denoted $\Delta(b)$. Note that one can generate multiple sets $\Delta(b)$, thus, following computations will depend on the particular instance of $\Delta(b)$.

Similarly, we could make the formation of $\Delta(\neg b)$, where the original attribute was replaced by its complement $\neg b$.

Having fixed $\Delta(b)$ and $\Delta(\neg b)$, we then used the approach described in \cite{ANO15} to rank the attributes by the \emph{relevance} parameter. 



For any attribute $a$, the relevance $rel_b(a)$ of $a$ to $b$ is computed based on frequency of $a$ appearing in the antecedents of implications/association rules related to $b$  in $b$-sectors $\Delta (b)$ and $\Delta (\neg b)$. From this definition one can see some relation between the relevance parameter and  \emph{conviction}, see for example  \cite{P17}. 

The computation of this parameter takes into account the support of each individual implication in the basis where $a$ appears. Since this time we have association rules of different confidence, we include the confidence into computation of the relevance as well.
For rule $\alpha=(X\rightarrow b)$, $conf(\alpha) = \frac{sup(X\cup b)}{sup(X)}$.

We believe that, for each attribute $a \in A\setminus{b}$, the important parameter of relevance of this attribute to $b \in A$ is a parameter of \emph{total support}, computed with respect to set of rules $\Delta$: 
\[
tsup_b(a)=\Sigma \{\frac{|sup(X)|}{|X|}\cdot conf(X\rightarrow b):  a\in X, (X\rightarrow b)\in \Delta (b)\} .
\]

Thus $tsup_b(a)$ shows the frequency of parameter $a$ appearing together with some other attributes in implications $X\rightarrow b$ of set $\Delta (b)$. The contribution of each implication $X\rightarrow b$, where $a \in X$, into the computation of total support of $a$ is higher when the support of $X$ is higher, i.e., column $a$ is marked by $1$ in more rows of the table, together with other attributes from $X$, but also when $X$ has fewer other attributes besides $a$.

While the frequent appearance of a particular attribute $a$ in implications $X\rightarrow b$ might indicate the relevance of $a$ to $b$, the same attribute may appear in implications $X\rightarrow \neg b$. 



Replacing $\Delta (b)$ by $\Delta (\neg b)$ in above formula, we can also compute the \emph{total support} of $\neg b$, for each $a \in A\setminus b$:
\[
tsup_{\neg b}(a)=\Sigma \{\frac{|sup(X)|}{|X|}\cdot conf(X\rightarrow \neg b) :  a\in X, (X\rightarrow \neg b)\in \Delta(\neg b)\} .
\]

Define now the relevance of parameter $a \in A\setminus b$ to parameter $b$, with respect to bases $\Delta (b)$ and $\Delta (\neg b)$: 
\[
rel_b(a) = \frac{tsup_b (a)}{tsup_{\neg b} (a) +1}.
\]

The highest relevance of $a$ is achieved by a combination of high total support of $a$ in rules $X\rightarrow b$ and low total support in rules $X\rightarrow \neg b$.
This parameter provides the ranking of all parameters $a \in A\setminus b$.

As we indicated above, the computation of the relevance can be done not only with implications but with any set of association rules $\Delta$. We believe that association rules of high confidence may provide a better set for computation of the relevance.

Observation with the data shows, and theoretical results confirm \cite{AN18}, that a rule $A\rightarrow b$ that fails in one or a few rows of table may appear through the set of implications $A\cup d \rightarrow b$, with multiple attributes $d$, which may inflate $tsup_b(a)$ for element $a \in A$. When one or a few rows failing the rule $A\rightarrow b$ are deleted, then $A \rightarrow b$ will be discovered, and the process of the aggregation will eliminate all the rules $A\cup d \rightarrow b$ from the final set of rules used for computation of the relevance.

\section{Technical implementation}

\pagestyle{empty}


Sequential algorithms are of little practical use when dealing with sufficiently computational complex problems and/or sufficiently large problems \cite{BPS2006}. Since the beginning of the demise of Dennard's scaling in the mid 2000's \cite{BE11}, the focus of mainstream computing hardware has moved away from sequential acceleration into parallel acceleration using multiple cores of execution \cite{KIMP2011}. Since then, algorithms can no longer rely on regular incremental improvements in serial execution speed due to Dennard's scaling. Instead the focus of high performance programming shifted to parallel algorithms. In addition, due to the demands of the age of big data and the decline in custom computing hardware, many of the algorithms in use today must also scale beyond the confines of a single homogeneous computing environment into a distributed heterogeneous execution environment \cite{BPS2006}. 

Finding association rules is a computationally complex problem \cite{AN15} and our approach is meant to address this issue using a collection of parallelizable algorithms that are scalable in a distributed processing environment.

\subsection{Algorithm Descriptions} 

At the heart of our new approach for discovering new implications are two highly parallelizable algorithms: 
\begin{enumerate}
	\item One Row Delete Algorithm (ORD) - Remove rows from the original table one row at a time and check for new discoverable rules.
	\item Multiple Row Delete Algorithm (MRD) - Remove groups of rows from the original table, aggregate the new rules and discover new rules that emerge from the aggregation. 
\end{enumerate}

The following code listings are written in pseudo C++ code; some code is omitted for brevity.


\lstset{breaklines=true}
\begin{lstlisting}[caption= {One Row Delete Algorithm(ORD)},label={lst:clist1}]
Table originalTable = {...}; // original table we are working on
ImplicationList impBaseList = generateBaseListOfImplications(originalTable);
DeleteRowList rowDelList = generateListOfRowsToDelete();
for(Row row2Del : rowDelList) { // for each row in row list
Table mutedTbl = createMutedTable(originalTable,row2Del);
ImplicationList impNewList = generateListOfImplications(mutedTbl);
impNewList = impNewList - impBaseList; // remove duplicates
calculateSupport(impNewList); //calculate support and report new implications 
}
\end{lstlisting}

The ORD algorithm (seen in listing \ref{lst:clist1}) breaks the problem of finding blocking rules in the original table into a list of $N$ sub-problems where $N$ is equal to the number of rows in the original table. In each of the sub problems we then need to discover new rules given a mutated original table with one of the N rows removed. 

We start by generating the base list of implications without changes to the original table (line 2).  Then we mutate the table (line 5) by removing different rows in each iteration of the loop and generate a new list of rules (line 6).  In the next step we make sure the new rules are previously undiscovered by comparing them to the original rules and removing any repetitions (line 7). In the last step (line 8) we report the new rules.  As can be seen in lines 5--8 of the algorithm, each iteration of the \emph{for} loop can be executed in parallel, therefore allowing us to spawn up to $N$ parallel executions of the DBasis algorithm.

Communication overhead and synchronization between parallel execution units are important factors in performance of parallel algorithms \cite{BPS2006}. In the ORD algorithm only the original table and the list of rows to be removed need to be communicated across parallel execution unit boundaries. In addition, no synchronization is necessary since each parallel unit can have its own copy of the original table and the row/rows to remove. 

\begin{lstlisting}[caption= {Multiple Row Delete Algorithm(MRD)},label={lst:clist2}]
Table originalTable = {...}; // original table we are working on
ImplicationList impBaseList = generateBaseListOfImplications(originalTable);
ImplicationList impAggregatedList = {};
DeleteRowCombList rowCombDelList = generateListOfCombOfRowsToDelete();
for(RowList rowComb2Del : rowCombDelList) { // for each row combination in row comb list
Table mutedTbl = createMutedTable(originalTable,rowComb2Del);
ImplicationList impNewList = generateListOfImplications(mutedTbl);
impNewList = impNewList - impAggregatedList; // remove duplicates
impAggregatedList = impAggregatedList + impNewList; // aggregate rules
}
calculateSupport(impNewList); //calculate support and report new implications 
\end{lstlisting}
 
The MRD algorithm (seen in listing \ref{lst:clist2}) shares much of the ideas and code with the ORD algorithm but differs in the following ways:
\begin{enumerate}
	\item It works on groups of rows instead of one group at a time (line 6)
	\item The new rules are aggregated inside the for loop (line 9)
	\item we only calculate the support in the end of the loop after aggregating all the new rules found in each loop iteration (line 11)
\end{enumerate}

The MRD algorithm allows us to execute lines 6--7 of each iteration of the for loop in parallel. Discovering new rules and accumulating the results of each iteration needs only to wait for individual iterations to complete in order to perform partial summing of the results (lines 8--9). The only part that needs to wait for all parallel iterations to complete is calculating the aggregated support (line 11).

\section{Testing}

\pagestyle{empty}

In addition to core programs discussed in previous section, this project develops a series of additional subroutines that perform secondary analysis of retrieved rules. We treat the output as an aggregated set, or basis of rules, that allows for statistical analysis.

First, we performed a series of tests with random matrices whose sizes mimicked real data at various densities (total number of ones in the matrix divided by the total number of entries) to investigate two things: the densities for which the algorithm may uncover a considerable number of rules, and to find the average relevance of rules in random matrices. Initial results indicate that the probability of obtaining high relevance of one attribute with respect to another remains very low for densities 0.3-0.4, and it increases when the density increases.

Testing shows that randomly generated binary tables may have pairs of attributes $a,b$ such that the relevance parameter $rel_b(a)$ could be considerably higher than 1. Several thousand random binary tables of fixed size $20\times 32$ and two different densities were analyzed, and their relevance characteristics are described in \emph{table 1}. A slightly more thorough comparison is done in $Fig. 1$, which shows the average relevance of random attributes. This data implies that a minimum relevance threshold might be recommended for practical data analysis given the possibility of unimportant attributes being ranked highly.

\begin{figure}
	\includegraphics[width=\linewidth]{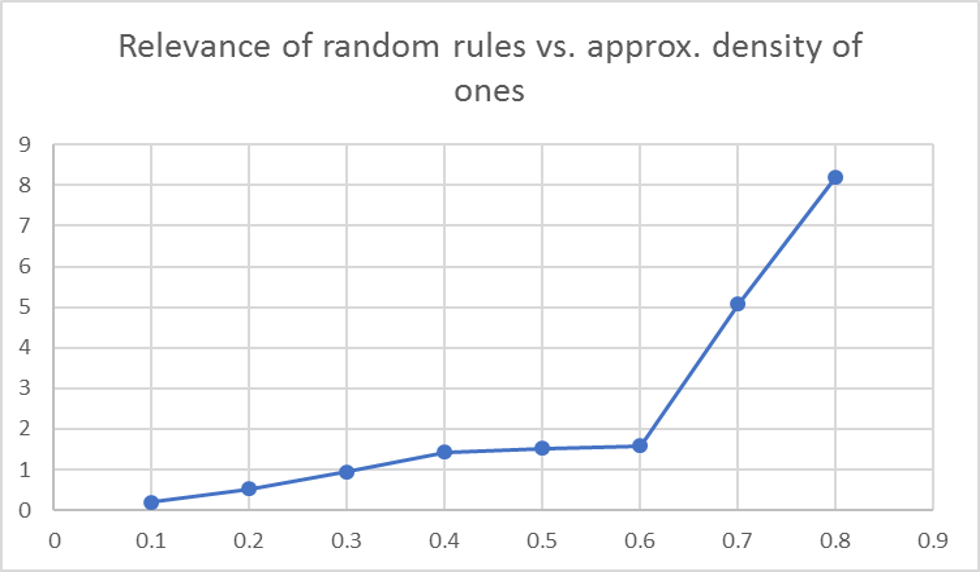}
	\caption{$x$-axis represents the density of ones in a binary table}
	\label{fig:graph of random relevance}
\end{figure}


\begin{table}[ht]
	\caption{Example of the relevance at two densities}
	\centering
	\setlength{\tabcolsep}{4pt}
	\begin{tabular}{c c c c c c c}
		\hline \hline
		Density & Min & Max & Average & 50th percentile & 75th percentile & 90th percentile \\ [.5ex]
		.3 & 0 & 19.75 & .951 & .520 & 1.182 & 2.212 \\
		.5 & 0 & 121.833 & 1.523 & .972 & 1.476 & 2.718 \\
	\end{tabular}
\label{table:nonlin}
\end{table}

We also simulated data that carries a few essential rules and imposed various levels of the noise to observe whether ORD or MRD would recapture the rules which were blocked by simulated noise, defined as a certain probability $p$ that $b$ would become $\neg b$. 
In many cases the data was recaptured, although the supporting statistics were diminished. Relevance decreased inversely proportional to the amount of noise added, 
leveling off at around 40-50\% noise.
After this, many attributes' relevance statistics were indistinguishable from the rules generated by noise.

These studies will be presented in the full size publication when all the testing is summarized and analyzed. We are planning to use the method for analysis of medical or biological data where we are looking for the rules of confidence close to 1.

For the purposes of this short presentation we ran a comparison of our approach with existing implementations of Apriori algorithm.

We conducted our tests on Ubuntu 14.04.1(VM) running on Intel(R) Xeon(R) CPU E5-2660 v2 @ 2.20GHz with 4 cores and 16GiB of DDR3 RAM provisioned.

Two data sets were used: one transactional set and one medical.

The first data set was taken from the Frequent Itemset Mining Dataset Repository \cite{FDS}. It is the retail market basket data from an anonymous Belgian retail store. We took the first 90 rows, converted them to a binary matrix format with size 90$\times$502 and (low) density 0.0162.

For our initial $D$-basis run, the time was 17.60 seconds, resulting in 422,273 implications of support $\geq 1$.

At the time of writing, our implementation of the MRD algorithm is sequential. It takes 42.63 minutes to run the $D$-basis program on the original table and then a batch of 200 smaller sub-tables (with several of rows removed).


For the Apriori implementation, we initially used Microsoft SQL Server Business Intelligence Development Studio, 2008 (Data Mining Technique – Microsoft Association Rules) \cite{ZL05}. It is worth noting that Apriori was designed specifically for mining association rules in retail data, which is normally produces the table of low density.
The parameters for Microsoft were: minimal support = 3, confidence (probability)= 1.0, max number of rules = 9000, the ranking by `importance' (= `lift'). 

Most of the 9000 rules were weaker rules than those given by the $D$-basis algorithm. For example the top rule found by Microsoft, ranking by the lift, was $96, 48 \rightarrow 95$, which corresponds to equivalence $96\leftrightarrow 95$ in the $D$-basis. On the other hand, some of the rules of support 5 and 8 from the $D$-basis did not appear on the 9000 list (majority of the rules in the 9000 set have support 1).

We then tested with the Apriori algorithm implemented in the R package 'arules' \cite{HBGK18}. We ran with the following parameters: minimum support = 3, max time = 10 (default), confidence (probability) = 0.8. There were 9 rules reported back from 'arules', with only two rules having confidence $< 1$.

The test with R found similar set of rules as the batch of 200 runs in MRD algorithm with 10 random rows removed at a time. All the rules found in R's Apriori function were  accounted for in the output of our program. On the other hand, our batch revealed that some of rules found by our algorithm are shorter versions of those found by the R algorithm. For example, our algorithm did not report the rule  $36, 48 \rightarrow 38$, because it reported rule $36 \rightarrow 38$, which is a shorter version of the rule.

Runtimes are immensely different, however, when one approaches non-sparse medical data such as the gene/survival data that was tested in \cite{ANO15} with the $D$-basis algorithm.

We re-tested again the data set treated in \cite{ANO15}, with 291 ovarian cancer patients, split into 4 survival groups, together with expression levels of 40 genes identified as essential for this type of cancer by other methods. The goal of analysis is to find a small group of genes, say 5 to 7, that may predict the good or bad survival of a patient. The data was converted to the binary format, where the first 80 columns represented expression levels of 40 genes, while columns $81$ through $84$ marked the survival subgroups of patients.
The ranking of 80 columns by the relevance parameter to any column $b =81,82,83,84$ would provide important information for medical specialists, and dependence of survival on identified subgroup of genes could be verified by other means such as Kaplan-Meier test \cite{KM58}. 

The $D$-basis algorithm allows to compute only a sector of the basis, with all the rules $X \rightarrow b$ with particular consequent $b$. We set $b=81$ and computed the $b$-sector of the $D$-basis in 135 seconds:
either 200,000 implications (rules of confidence = 1) of minimum support 3, or just 91 of minimum support 5. 


For the testing of MRD algorithm,
we ran a batch of 25 runs of shorter tables removing randomly 20 rows at a time, requesting the rules $X\rightarrow 81$ of minimum support 5. 

Sequential time was 51.11 minutes, with roughly 125 seconds per run. 
The average confidence of the new rules was 0.93, and a total of 219 rules were retrieved.


Thus, together with 91 rules of confidence = 1, we found the set $\Delta (b)$ of 310 rules of high confidence. Note that no aggregation or relevance computation was performed, because the purpose of the test was the comparison with software that does the retrieval of the rules, but not the ranking of the attributes.

Note that the density of the medical data matrix is 0.34 (compared to just 0.0162 in transactional data), and the rules have a tendency to be long.
Among the rules of confidence = 1 there were 35 rules $X\rightarrow b$ with $|X|=6$ and 7 rules with $|X|=7$. Among rules of less confidence, there were numerous rules with $|X| >9$.


When requesting 'arules' in R, the program runs with the user's parameters. In order to reduce the time taken for the computation, there is a parameter \emph{maxtime}, which limits the time used per frequent sets of attributes.  This parameter stops the program when the time to produce some set of rules per subset exceeds expected times of computation. 

For Apriori to analyze a more dense data set such as the medical data set, it will need to generate all frequent sets containing $b= 81,82,83,84$, but the number of  such frequent sets will be much larger than the eventual rules with consequent $b$. This highlights exponential time complexity of Apriori vs. the sub-exponential time it would take to analyze the same data set with the $D$-basis algorithm.

Requesting `arules' on the equivalent data for the data set in \cite{ANO15} was unsuccessful due to the inordinate memory complexity and time complexity of the algorithm. We had successfully analyzed subsets of length 7 before the memory demands were too high. On the same machine that ran the $D$-basis program, Apriori in R ran with minimum support = 5 and maximum length of rules = 7 and took 43.67 seconds. 

While faster than the $D$-basis program, the set of rules was restricted, and any attempt to test with larger parameters stopped execution due to large memory requirements. These rules were also limited in size so that only rules with $|X| \leq 6$ were generated.


\section{Conclusion and Future Work}
In this paper we have discussed the development of an algorithm for the retrieval of association rules in a binary table i.e., a table consisting of ones and zeroes as another representation of the data. This is done via dualizing the hypergraph associated with the dataset, then reducing the task of rule generation to traversing this associated hypergraph via any sub-exponential time-complexity algorithm \cite{AN15}.

Several development proceedings were discussed, including the eventual parallelization of the program and a ``top-down" method of retrieving rules which hold in all rows of the table except a few removed rows. Analyzing a slightly smaller sub-table allows to discover rules which have high confidence and may fail in a few rows due to noise in the data. The ability for this code to be parallelized and its low theoretical time complexity make it a powerful tool for data mining.

The \textit{relevance} parameter for determining the importance of one attribute to an outcome was also tested to see how it might be used in analyzing real data. It was seen that even random rules could produce a notable relevance values, summarized in \textit{Fig. 1}. Then, a synthetic rule was constructed and noise was added to the matrix in order to see what effect noise would have on the relevance of the parameters of the constructed rule, and whether or not the process of retrieving rules of high confidence via row deletion could recover the synthetic rule if it was blocked. The first part of these tests revealed that relevance stayed relatively stable up until approximately 30-40\% noise, and that depending on the noise and how many rows were removed, the process could recover the synthetic rule albeit with less support and confidence.

Lastly, we tested several methods for ``ranking" which rows should be deleted in order to recover rules lost to noise. 
One of tested heuristic is based upon analyzing the inverted table, however the reason for this heuristic's efficacy is not yet identified.

The development of this program is still underway, with real implementation of distributed computing expected later in 2018. We currently have several data sets in biology, medicine and meteorology which we plan to explore, working in collaboration with Biology Department, Geology, Environment and Sustainability Department of Hofstra University,as well as Donald and Barbara Zucker School of Hofstra-Northwell. We also plan to continue the collaboration with the Cancer Center of University of Hawai'i, and contribute to the exploration of data sets of various cancers, which combines several available methods \cite{N18}.\\

{\bf Acknowledgements.} We used for testing the data set on ovarian cancer available at \cite{TCGA}, which was also used in \cite{ANO15} and managed in the lab of Dr.~Gordon Okimoto, at Cancer Center of Univeristy Hawai`i. We thank the support of Computer Science Department of Hofstra University for providing the virtual machine environment for this project. The second author was partly supported by Honor's College of Hofstra University's Research Assistant Grant.


\begin{thebibliography}{1}

\bibitem{AN15} K. Adaricheva, and J.B. Nation, \emph{Discovery of the $D$-basis in binary tables based on hypergraph dualization}, v.658 (2017), Theoretical Computer Science, Part B, 307--315.

\bibitem{ANO15} K.~Adaricheva, J.B.~Nation, G.~Okimoto, V.~Adarichev, A.~Amanbekkyzy, S.~Sarkar, A.~Sailanbayev, N.~Seidalin, and K.~Alibek, \emph{Measuring the Implications of the $D$-basis in Analysis of Data in Biomedical Studies}, Proceedings of ICFCA-15, Nerja, Spain; Springer, 2015,  39--57.


\bibitem{ANR11} K.~Adaricheva, J.B.~Nation and R.~Rand,  \emph{Ordered direct implicational basis of a finite closure system}, Disc. Appl. Math. \textbf{161} (2013),  707--723.


\bibitem{AN18} K.Adaricheva and T. Ninesling, \emph{Direct and binary-direct bases for one-set updates of a closure system}, manuscript; presented in poster session of ICFCA-2018 http://icfca2017.irisa.fr/program/accepted-papers/


\bibitem{A96} R.~Agrawal, H.~Mannila, R.~Srikant, H.~Toivonen and A.I.~Verkamo, \emph{Fast discovery of association rules}, Advances in Knowledge discovery and data mining, AAAI Press, Menlo Park, California (1996), 307--328.



\bibitem{BPBBK16} Guillaume Bosc, Marc Plantevit, Jean-FranÃ§ois Boulicaut, Moustafa Bensafi, and Mehdi
Kaytoue, h (odor): Interactive discovery of hypotheses on the structure-odor relationship in neuroscience, in ECML/PKDD 2016 (Demo), 2016.


\bibitem{B10} J.L. Balc\'azar, \emph{Redundancy, deduction schemes, and minimum-size bases for association rules}, Log. Meth. Comput. Sci. \textbf{6} (2010), 2:3, 1--33.

\bibitem{BPS2006} Gordon Bell, Jim Gray, and Alex Szalay. \newblock Petascale computational systems.\newblock {\em Computer}, 39(1):110--112, 2006.



\bibitem{BE11} Raffaele Bolla, Roberto Bruschi, Franco Davoli, and Flavio Cucchietti. \newblock Energy efficiency in the future internet: A survey of existing approaches and trends in energy-aware fixed network infrastructures.
\newblock {\em IEEE Communications Surveys \& Tutorials}, 13(2):223--244, 2011.




\bibitem{FK96} M.~Fredman and L.~Khachiyan, \emph{On the complexity of dualization of monotone disjunctive normal forms}, J. Algorithms \textbf{21} (1996), 618--628.

\bibitem{GW99} B. Ganter, and R.Wille, Formal concept Analysis: Mathematical Foundations, Springer, 1999.


\bibitem{HBGK18} M.~Hahsler, C.~Buchta,  G.~Bettina Gruen and K.~Hornik, \emph{arules: Mining Association Rules and Frequent Itemsets}, R package version 1.6-0. https://CRAN.R-project.org/package=arules
\textbf{2018}

\bibitem{KM58} E.L.~Kaplan and P.~Meier, \emph{Nonparametric estimation from incomplete observations}, J. Amer. Statist. Assn. \textbf{53} N282 (1958), 457--481.

\bibitem{KIMP2011}
Jonathan Koomey, Stephen Berard, Marla Sanchez, and Henry Wong.
\newblock Implications of historical trends in the electrical efficiency of
computing.
\newblock {\em IEEE Annals of the History of Computing}, 33(3):46--54, 2011.


\bibitem{K02} M.~Kryszkiewicz, \emph{Concise representation of association rules}, Proceedings of the ESF Exploratory Workshop on Pattern Detection and Discovery, Springer-Verlag, London, UK, 92--109.

\bibitem{MU13} K.~Murakami and T.~Uno, \emph{Efficient algorithms for dualizing large scale hypergraphs}, Disc. Appl. Math. \textbf{170} (2014), 83--94. 

\bibitem{N18} J. B. Nation, G. Okimoto, T. Wenska, A. Achari,
J. Maligro, T. Yoshioka, and E. Zitello, A Comparative analysis of MRNA 
expression for sixteen different cancers, preprint, $http://www.math.hawaii.edu\/~jb/lust\_2017\_615.pdf$


\bibitem{SC09} Arnaud Soulet and Bruno CrÃ©milleux, \emph{
Mining constraint-based patterns using automatic relaxation},
Intell. Data Anal., 13(1):109--133, 2009.

\bibitem{SRPC11} Arnaud Soulet, Chedy RaÃ¯ssi, Marc Plantevit, and Bruno Cremilleux,
\emph{Mining dominant patterns in the sky}
In IEEE 11th Int. Conf on Data Mining (ICDM 2011), pages 655--664. IEEE, 2011.


\bibitem{S04} C.~Spearman, \emph{The proof and measurement of association between two things}, Amer. J. Psychol. \textbf{15} (1904), 72--101.


\bibitem{P17} D. Prajapati, S. Garg and N.C.Chanhan \emph{Interesting association rule mining with consistent and inconsistent rule detection from big sales data in distributed environment}, Future Computing and Informatics Journal \textbf{2} (2017), 19--30.

\bibitem{ZL05} Zhao Hui Tang, Jamie MacLennan, Data Mining with SQL server, Wiley 2005.

\bibitem{TCGA}
The Cancer Genome Atlas Research Network, \emph{The Cancer Genome Atlas Pan-Cancer analysis project}, Nature Genetics ~\textbf{45} (2013), 1113--1120.

\bibitem{R13}
R Core Team, \emph{R: A language and environment for statistical
  computing}, R Foundation for Statistical Computing, Vienna, Austria, 2013.
  URL http://www.R-project.org/.

\bibitem{FDS} Frequent Itemset Mining Dataset Repository, publicly available at http:// fimi. ua. ac.be /data/retail.dat

\end{thebibliography}
\end{document}